\documentclass[conference]{IEEEtran}
\usepackage{authblk}

\usepackage[T1]{fontenc}
\usepackage{graphicx}

\AtBeginDocument{%
  }

\usepackage{amsmath,amssymb}
\usepackage{algorithmic}
\usepackage{graphicx}
\usepackage{bmpsize}
\usepackage{xcolor}
\usepackage{lipsum}
\usepackage{cleveref}

\usepackage{tabularx}

\usepackage{siunitx}

\usepackage{pifont}

\usepackage{comment}

\usepackage[utf8]{inputenc}

\usepackage[T1]{fontenc}
\usepackage{url}                
\usepackage{xcolor}             

\usepackage{tikz}
\usepackage{rotating}
\usepackage{diagbox}

\usepackage{listings}
\usepackage{setspace}

\usepackage{multirow}
\usepackage{longtable}

\usepackage[flushleft]{threeparttable}
\usepackage{pdfpages}
\usepackage{enumitem}
\usepackage{placeins}
\usepackage{subcaption}
\usepackage{cmtt}
\usepackage{bold-extra}
\usepackage{soul}
\setuldepth{lifter}

\newcommand{\sys}{\mbox{\textsc{D-LiFT}}\xspace}
\newcommand{\syss}{\mbox{\textsc{D-Score}}\xspace}

\newcommand{\AK}[1]{\textcolor{red}{AK: #1}}
\usepackage{ifthen}
\newcommand{\switchsome}[1]{%
  \ifthenelse{\equal{#1}{0}}{\renewcommand{\AK}[1]{}\renewcommand{\MZ}[1]{}}{}
  }

\usepackage[export]{adjustbox}
\usepackage[ruled, lined, linesnumbered, commentsnumbered, longend]{algorithm2e}
\usepackage{pdfrender}

\def\thickhline{\noalign{\hrule height 1.1pt}}

\makeatletter
\usepackage{tikz}
\usetikzlibrary{calc}
\newcommand*\circled[1]{\tikz[baseline=(char.base)]{
    \node[shape=circle, draw, inner sep=0.5pt,
        minimum height={\f@size},] (char) {\vphantom{WAH1g}#1};}}
\makeatother

\definecolor{mGreen}{rgb}{0,0.6,0}
\definecolor{mGray}{rgb}{0.5,0.5,0.5}
\definecolor{mPurple}{rgb}{0.58,0,0.82}
\definecolor{backgroundColour}{rgb}{0.95,0.95,0.92}
\definecolor{comment}{RGB}{0,128,0}     
\definecolor{string}{RGB}{255,0,0}      
\definecolor{instruction}{RGB}{0,0,255} 
\definecolor{directive}{RGB}{128,0,128} 
\definecolor{register}{RGB}{128,0,0}    

\lstdefinestyle{cstyle}
{
 numbers=left,
    stepnumber=1,
        breaklines=true,
    breakatwhitespace=true,
  frame=leftline,
  captionpos=t,
  xleftmargin=0.01em,
  xrightmargin=0.01em,
  language=C,
  numbersep=5pt,
  showstringspaces=false,
  keywordstyle=\color{black},
  commentstyle=\color{comment},
  identifierstyle=\color{black},
  stringstyle=\color{orange},
  basicstyle=\fontfamily{cmtt}\tiny,
  numberstyle=\tiny,
  escapeinside={(@}{@)},
  moredelim=**[is][\color{green!70!black}\bfseries]{§}{§},
}
\usepackage{color, colortbl}
\usepackage{float}

\definecolor{Gray}{gray}{0.9}
\newsavebox{\verbsavebox}

\begin{document}

\date{}

\title{Fine Tuning LLM-Based Decompiler Backend for Output Code Quality and Trustworthiness}


\author[1]{Muqi Zou}
\author[1]{Hongyu Cai}
\author[1]{Hongwei Wu}
\author[2]{Zion Leonahenahe Basque}
\author[3]{Arslan Khan}
\author[1]{\\Z. Berkay Celik}
\author[1]{Dave (Jing) Tian}
\author[1]{Antonio Bianchi}
\author[2]{Ruoyu (Fish) Wang}
\author[1]{Dongyan Xu}
\affil[1]{Purdue University}
\affil[2]{Arizona State University}
\affil[3]{Pennsylvania State University}
\affil[1]{\textit {\{zou116, hongyu, wu1685, zcelik, daveti, antoniob, dxu\}@purdue.edu}}
\affil[2]{\textit{\{zbasque,fishw\}@asu.edu}}
\affil[3]{\textit{arslankhan@psu.edu}}    
\maketitle

\begin{abstract}
As a prevalent function in many software security tasks, decompilation reconstructs human-readable source code from binaries.
Despite recent advances, decompiler outputs remain difficult to read.
With the advent of large language models (LLMs), researchers recently began to explore the potential of LLMs to improve decompiler output.
Nevertheless, our study of these approaches reveals their problems, such as introducing new syntactic and semantic errors and relying on unreliable accuracy validation, making the decompiled code less trustworthy.
In this paper, we present \sys, an enhanced ``decompiler-LLM'' pipeline with a fine-tuned LLM using {\em code quality-aware} reinforcement learning.
Unlike prior work that overlooks preserving accuracy, \sys adheres to a key principle for enhancing the quality and trustworthiness of decompiled code: \textit{preserving accuracy while improving readability}.
Central to \sys, we propose \syss, an integrated code quality assessment system to score the decompiled source code from multiple aspects, and use it to guide reinforcement learning-based LLM fine-tuning and to select the best output during inference.
In line with our principle, \syss assigns low scores to any inaccurate output and only awards higher scores to code that passes the accuracy check. 
Our implementation, based on Ghidra and a range of LLMs, demonstrates significant improvements for the accurately decompiled code from the coreutils and util-linux projects.
Compared to baseline LLMs without \syss-driven fine-tuning, our trained LLMs produce 55.3\% more improved decompiled functions.
Overall, \sys improves the quality of 68.2\% of all the functions produced by the native decompiler.

\end{abstract}
\section{Introduction}

Binary decompilation, as an important program analysis function, is the process of generating high-level human-comprehensible source code from a binary program. Decompilation supports numerous software management and security tasks, such as reverse engineering~\cite{observation2020Daniel}, vulnerability discovery~\cite{convergence2022Mantovani,QueryX2023han}, program patching and hardening~\cite{Pemma2022automatically}, and cyber forensics~\cite{decomperson}. 
However, source code generated by decompilers is often hard for human users to read~\cite{Zion2024ahoy,zhuo2021osprey,qibin2022augmenting,Lin2010AutomaticRE,chang2024tygr,Dream,Dreamplusplus,brumley2013native,TIE}.
Moreover, decompiled outputs often suffer from syntactic and semantic correctness issues~\cite{zhibo2020howfar,Muqi2024Dhelix,cao2024evaluating,Taxonomy2024luke,zaobin2wring2025, GorzynskiECOOP2025}:

A piece of decompiled code may not compile or preserve semantics of the original binary, thereby undermining the trustworthiness of the decompiled output.

In recent years, with the advent of Large Language Models (LLMs), researchers have started to explore the potential of LLMs to generate and improve decompiled code~\cite{danning2024resym,Tan2024LLM4DecompileDB,Hu2024DeGPTOD,Katz2019TowardsND,fu2019coda,linxin2025beyond,Wong2023RefiningDC,xiangzhe2025unleash,lukefast2026,idioms2026luke}.
For example, to improve the readability of decompiled outputs, 
LLM4Decompile~\cite{Tan2024LLM4DecompileDB} trains new LLMs via supervised fine-tuning (SFT), and DeGPT~\cite{Hu2024DeGPTOD} introduces a three-role mechanism to help verify LLMs' output during inference.
These LLM-assisted decompilation improvement tools have rapidly gained attention.
For instance, LLM4Decompile was ranked 4th in GitHub Daily Trending.~\looseness=-1

Unfortunately, applying LLMs to improve decompiled code may introduce inaccuracy, further undermining its trustworthiness.
For instance, \Cref{fig:introcode} in the Appendix shows how GPT-o4-mini~\cite{openai2024gpt4ocard} can easily introduce semantic errors. 
Furthermore, our analysis of the above state-of-the-art LLM-based methods for decompiled code improvement also shows significant limitations in preserving the original semantics.
For LLM4Decompile, our study (detailed in \Cref{baselineselect}) reveals that, even when the input is error-free, new errors are found in 93.2\% of the functions generated by its LLM.
For DeGPT, it cannot enhance the LLM's capability to generate more accurate decompiled code, since it only provides a non-deterministic semantic check after inference.
These findings show a clear need for a more rigorous, LLM-driven decompiled-code enhancement pipeline, one that incorporates a novel training paradigm to produce smarter models and a rigorous inference-time accuracy check to ensure high-readability and semantics-preserving outputs.
 
Hence, we propose \sys \footnote{``\sys'' reflects the decompilation pipeline with a ``\textbf{D}ecompiler'' front-end and an ``\textbf{L}LM w\textbf{i}th \textbf{F}ine \textbf{T}uning'' back-end.}, an enhanced decompiler-LLM pipeline with a fine-tuned LLM using code quality-aware reinforcement learning. 
\sys takes a baseline LLM, a decompiler, and a set of real-world binaries as inputs, and outputs decompiled code of higher quality using an enhanced decompiler-LLM pipeline.
To this end, \sys first leverages reinforcement learning (RL) to fine-tune the baseline LLM and then, at inference, selects the top-scored output from the baseline LLM, the fine-tuned LLM, or the native decompiler.

Central to \sys is \syss, a novel code quality and trustworthiness assessment function, which provides multi-aspect scoring for LLM-generated decompiled code based on accuracy and readability.
Unlike prior work, \syss follows a key principle: {\em preserving accuracy while improving readability}. 
Specifically, first, to assess the accuracy of decompiled code, \syss employs a compiler to generate the {\em syntax} feedback and symbolic execution to compare the {\em semantics} of the generated code against the corresponding function in the original binary.
After that, to assess the readability of decompiled code, \syss computes a score by applying two established readability metrics to compare the LLM's output with the original decompiled code.
By using \syss as the reward function in reinforcement learning, \sys enhances the LLMs' ability to generate higher-quality decompiled code.
Meanwhile, by using \syss at inference, \sys comparatively selects the best result among the native decompiler, the baseline LLM, and the fine-tuned LLM.

We implement \sys using Ghidra~\cite{ghidra} and a number of LLMs, and evaluate \sys using functions from the coreutils~\cite{coreutils} and util-linux~\cite{util-linux} projects. 
As measured by \syss, models fine-tuned by \sys improve the quality of 55.3\% more functions compared to the baseline LLMs.
Interestingly, we observe a significant ``improve-ability gap'' between the accurate and inaccurate decompiled functions originally produced by the decompiler.
All LLMs, including baseline and fine-tuned models, face challenges in improving the originally inaccurate decompiled code, with improvement rates of just 8.02\% for the originally inaccurate functions, in contrast to 86.2\% for the originally accurate ones.
Overall, for functions that are accurately decompiled, \sys improves the quality (measured by \syss) of 68.2\% of them, where 47.3\% of them are improved by the \sys fine-tuned LLM, and only 20.9\% by the baseline LLM.
In summary, our main contributions include the following: 
\begin{itemize}
    \item We design \sys, an enhanced decompiler-LLM pipeline with a fine-tuned LLM using code quality-aware RL to improve the quality and trustworthiness of the decompiled code, adhering to the {\em preserving accuracy while improving readability}.
    \item We propose \syss, an integrated scoring scheme to quantitatively assess the quality of decompiled code. 
    The scheme incorporates existing code analysis techniques and established readability metrics to deliver a multi-aspect evaluation of decompiled code, assessing both syntactic and semantic correctness as well as readability properties to provide effective feedback to the LLM for training and a quantitative selection metric for inference.
    \item We implement \sys based on Ghidra and three LLMs, and achieve significant decompiled code improvement for the benchmark functions. 
\end{itemize}

\section{Motivation}

\label{motivation}
Though improving decompiled code via LLMs is prevalent, it has limitations. 

\textit{LLMs introduce inaccuracy during the decompiled code improvement, making the code less trustworthy}. 
LLMs, operating as probabilistic models that generate text through token-level predictions, systematically introduce inaccuracies in code quality improvement tasks, as evidenced by recent empirical studies documenting pervasive hallucination phenomena~\cite{llmhallucination2025zhang,Pan_2024,jiang2024surveylargelanguagemodels}.
In the decompiled-code improvement scenario, we also observe that LLM-refined output frequently shows new errors. 
Here, we follow the previous work~\cite{zhibo2020howfar} to define the inaccuracy in the decompiled code as either {\em syntax errors} that prevent successful compilation or {\em semantic deviations} between the LLM‑generated code and the original binary’s behavior.
As shown in \Cref{tab:baselineresult}, an average of 44.2\% of functions that were originally accurate become inaccurate after LLM processing. 
For the root cause, as case study examples shown later in \Cref{casestudy}, we observed that LLMs frequently make errors on small details, such as omitting brackets or instructions, or referencing wrong variables, that are hard to spot yet vital for accuracy.

These observations reveal that though LLMs may improve readability, they often introduce inaccuracy -- hence untrustworthiness -- of decompiled code, a factor that prior studies~\cite{Muqi2024Dhelix,zhibo2020howfar,cao2024evaluating,Taxonomy2024luke,yangbin2wrong} have identified as vital for effective decompilation.
Hence, we propose an overarching principle for improving decompiled code quality and trustworthiness: \textit{preserving accuracy while improving readability}. 

\textit{However, the existing "decompiler-LLM" pipeline does not follow the above principle.}
Specifically, LLM4Decompile, while effectively boosting LLM's capability of generating more readable code through training, it compromises the accuracy of its output, as shown in \Cref{tab:baselineresult}.
While running its \texttt{MSSC}, a mutated unit test, to check LLM output code semantics among inference, DeGPT uses random inputs that produce non-deterministic semantic check results and omits the check for syntax errors.
In light of these limitations, we are motivated to develop a more effective "decompiler-LLM" pipeline, where the LLM backend is specifically designed for improving the readability of the decompiled code while preserving its syntactic and semantic accuracy for trustworthiness.

\section{Design Challenges}
\label{designchallenge}


\subsection{Framework Design}
\label{designchallenge1}

\noindent\subsubsection{Training}
Improving the code quality of LLMs through training is not new~\cite{le2022coderl,shojaee2023execution,Tan2024LLM4DecompileDB,palit2024generating,dou2024stepcoder,lukefast2026}.
However, many of them are challenging to adapt to decompiled code improvement.
Unlike general code improvement tasks, enhancing decompiled code faces a unique challenge: the existence of multiple valid and trustworthy ground truths.
Specifically, a single binary may be compiled from multiple semantically equivalent source programs, each of which should be considered a valid ground truth. 
For instance, simple transformations such as variable renaming or converting a for loop into an equivalent while loop produce the same binary outputs while significantly increasing source-level edit distance.
In practice, however, researchers typically have access to only one of these source variants as the ground truth.
Moreover, many fine-tuning methods, such as supervised fine-tuning (SFT) and the actor-critic paradigm in RL, accept only a single reference to compute the training loss, overlooking the possible existence of a full set of correct alternatives.
This constraint may degrade a model’s effectiveness, as exemplified by LLM4Decompile’s use of SFT on one source code.

\noindent\subsubsection{Inference}
In scenarios where we have multiple enhanced code outputs, from the decompiler itself or various LLMs, an objective and quantitative metric is essential to rank quality.
Unfortunately, the only existing framework, DeGPT, offers no automated readability metric, relying on human participants' assessments.

\subsection{Assessment of Decompiled Code Quality}
\label{designchallenge2}
As noted above, the existence of multiple valid ground truths (during training) and enhanced candidates (at inference) needs a quantitative code-quality metric to evaluate each candidate. 
Nevertheless, to the best of our knowledge, \textit{NO} existing metric meets the dual requirement of maintaining accuracy and trustworthiness while enhancing readability.

\textit{For decompiler accuracy}, as introduced in \Cref{motivation}, two aspects of accuracy, syntactic and semantic, need to be evaluated.
For syntax validation, relying on standard compiler checks is a reliable choice.
However, there is no widely adopted standard for semantic validation between decompiled code and its corresponding binary.
Specifically, since the decompiled code assessment function must be deterministic, methods such as fuzzing~\cite{yangbin2wrong} and its derivatives, like \texttt{MSSC}~\cite{cao2024evaluating} from DeGPT, are excluded.
Additionally, to ensure the semantics are compared directly with the binary instead of the original source code, we exclude approaches that rely on source-dependent checks, such as unit tests or \texttt{Alive2}~\cite{alive22021lopes}.
That leaves \texttt{D-helix}~\cite{Muqi2024Dhelix}, which uses an iterative re-compiler to reveal syntax errors and symbolic execution to check the code's semantics against the original binary.
However, integrating \texttt{D-helix} directly into our framework introduces its own challenge.
Notably, \texttt{D-helix} still relies on certain decompiled code output artifacts, e.g., external function calls, to analyze.
Since LLM-generated code frequently omits instructions, these artifacts may be missing or unreliable, undermining \texttt{D-helix}’s effectiveness.
Furthermore, in the absence of the original source code, there is no authoritative ground truth for the number, types, or ordering of arguments of the external function.

\textit{For readability,} the most direct approach is to run a human study, e.g., pairwise comparisons on platforms such as CodeArena~\cite{du-etal-2025-codearena}, and then train a model on the collected preferences to score readability (an RLHF-style pipeline). 
However, such scoring models are expensive to build at scale: for instance, as shown in OpenAI’s study~\cite{openairlhf}, training a model for scoring the text summarization required nearly 64k human pairwise preferences and “thousands of labeler hours and significant researcher time”, illustrating the cost of large, high-quality preference datasets. 
Moreover, most human-comparison settings yield binary outcomes (0–1 preferences), whereas our RL framework requires scalar readability scores per candidate. 

Regarding human-validated, automatic readability metrics, most existing metrics~\cite{buse1010coderead,marvin2020emperread,posnet2011softwarread,scalab2018codereadability} ignore decompiler-specific artifacts (e.g., the number of references/dereferences).
To the best of our knowledge, R2I~\cite{eom2024r2i} is the only metric tailored specifically for decompiled code.
However, since it is a relative measure, whose features are derived to compare between multiple decompilers, it cannot assign an absolute score to a single piece of LLM‑generated code. 
For instance, the feature, ''number of unnecessary goto labels'', is calculated by contrasting angr~\cite{angr} or RetDec~\cite{RetDec} outputs against those from Ghidra~\cite{ghidra} and Hex‑Rays~\cite{hex}, which cannot be computed when only one decompilation exists.

In summary, our principle, preserving accuracy while improving readability, cannot be satisfied by any single code quality metric currently available.
\section{\sys Design}

\begin{figure*}[t]
  \centering
  \includegraphics[width=0.83\linewidth]{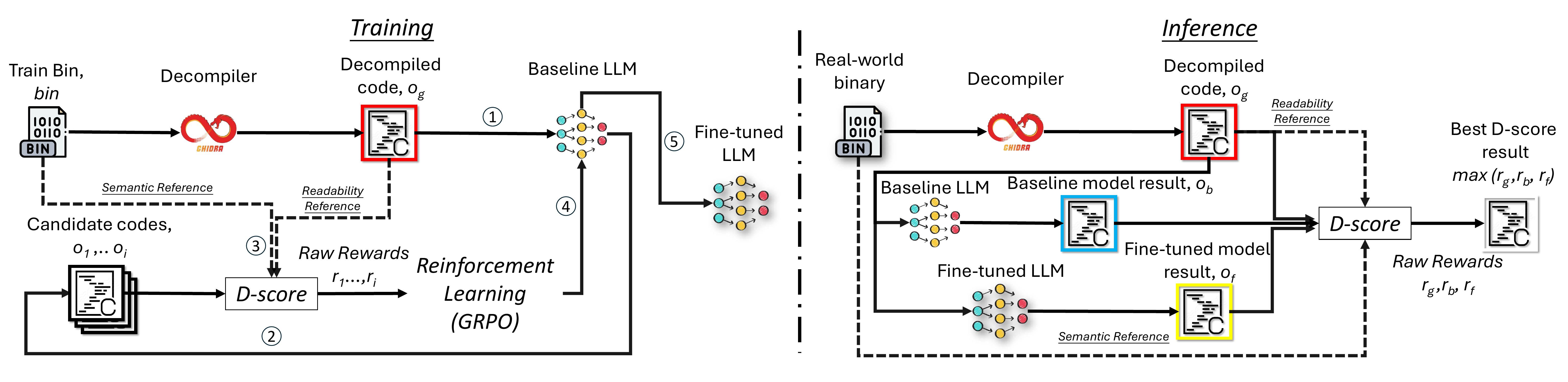}
  \caption{This figure shows how \sys fine-tunes the model and selects outputs.}
   \label{dliftframework}
\end{figure*}

\sys is an automatic decompilation pipeline that consists of a decompiler front-end and an LLM back-end capable of improving the quality and the trustworthiness of decompiled code. 
The workflow of \sys, illustrated in \Cref{dliftframework}, consists of a training phase (\Cref{RLdesign}) and an inference phase
(\Cref{bestpickdesign}).
Specifically, in the training phase, \sys takes three inputs: a baseline model (i.e., the LLM to train), a set of training binaries, and a decompiler.
It employs reinforcement learning to fine-tune the LLM, ultimately producing a model with improved decompilation capabilities.
Then, in the inference phase, \sys 
first uses the decompiler to generate the initial decompiled code for an input binary. It then applies both the baseline and the fine-tuned LLMs to refine the code and selects the best-quality result among all outputs.
To guide the training and inference phase 
, \sys introduces a multi-aspect code quality assessment metric, \syss, that evaluates a piece of candidate code with respect to both accuracy and readability (\Cref{RWdesign}).
Specifically, for accuracy, \syss adopts the stepwise approach: first verifying syntax (\Cref{syntaxdesign}), then validating semantics against the original binary (\Cref{semanticdesign}).
Only if a candidate passes these accuracy checks will \syss move on to assess readability (\Cref{readabilitydesign}).
~\looseness=-1

\subsection{System Workflow}
\sys works by first fine-tuning the LLM via GRPO to train the LLM with accuracy-aware rewards. Then at runtime, it will select the best-quality decompiled code for each input binary, among the outputs of the decompiler, the baseline LLM, and the fine-tuned LLM, by comparing their \syss.

\subsubsection{Training Phase}
\label{RLdesign}
The challenge described in \Cref{designchallenge1} highlights a critical limitation in current approaches: decompiled code can map to multiple semantically equivalent source representations, each representing a valid ground truth.  
This multiplicity poses significant training difficulties for supervised fine-tuning (SFT) approaches, such as LLM4decompile, which are designed to work with only one correct answer.
Although reinforcement learning generates multiple candidates during training, theoretically providing access to various valid solutions, conventional RL policies maintain the single ground truth constraint when computing the loss.
Fortunately, Group Relative Policy Optimization (GRPO)~\cite{shao2024deepseekmathpushinglimitsmathematical} overcomes this constraint by generating the loss from the normalized rewards across multiple candidate outputs.

\Cref{dliftframework} illustrates how \sys integrates GRPO into the training. 
\circled{1} Given a training binary, denoted as $bin$, \sys invokes the decompiler to produce an original decompiled output, denoted as $o_g$.
\circled{2} Using the $o_g$ as input, the baseline LLM generates a set of candidate refinements,$\{o_1,o_2,...o_i\}$.
\circled{3} For each candidate, $o_i$, \sys computes a reward score, $r_i$, using \syss (see \Cref{RWdesign} for details).
\circled{4} Next, GRPO normalizes these rewards, $\mathbf{r}$ = $\{r_1,r_2,...\}$ within each candidate group according to its standard formulation:  
\begin{equation}
\hat{A}_{i, t}=\widetilde{r}_i=\frac{r_i-\operatorname{mean}(\mathbf{r})}{\operatorname{std}(\mathbf{r})}
\end{equation}
and uses these normalized rewards to compute the loss for fine-tuning the model.
Given that the decompilation task accepts multiple valid ground truths, \sys provides no ground truth code; it relies exclusively on \syss-generated scores, removing any dependence on a particular reference solution, e.g., source code. 
Beyond reducing memory and computation, this choice aligns with recent results showing that reference-free training can be more effective~\cite{hu2025openreasonerzeroopensourceapproach,liu2025understandingr1zeroliketrainingcritical}.
\circled{5} Finally, once training iterations complete, \sys outputs the fully fine-tuned LLM.

\subsubsection{Inference Phase}
\label{bestpickdesign}
As described in \Cref{motivation}, LLMs often introduce errors when improving code, creating a need for a quantitative metric-guided selection system.
To address this, we designed the selection system
as shown in \Cref{dliftframework}.
This system works in two steps: 
\circled{1} It first takes a user's binary file and runs a decompiler to get the original code. It then feeds that code to both the baseline LLM and our fine-tuned LLM to generate two improved versions. 
\circled{2} By leveraging \syss, the same scoring tool applied during training, it evaluates all three outputs (the original, the baseline's, and the fine-tuned's), selecting the highest-scoring one as the final result for the user.

\subsection{\syss Design}
\label{RWdesign}
\begin{figure}[h]
\includegraphics[width=0.36\linewidth]{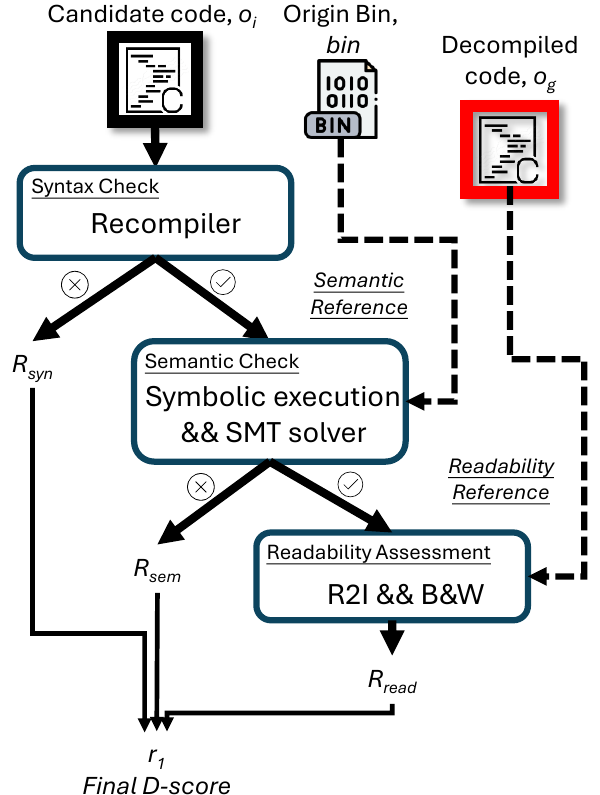}
  \centering
  \caption{\syss, following our principle of \textit{preserving accuracy while improving readability}.}
  \label{qscoreframework}
\end{figure}

\Cref{qscoreframework} shows the overall procedure of \syss. 
As a code quality/trustworthiness assessment function, \syss takes three inputs: a candidate code waiting for evaluation, $o_i$, its original binary, $bin$, and its original decompiled code from the decompiler, $o_g$, and outputs an absolute score $r_i$ reflecting both the accuracy and readability of $o_i$. 
Following our principle, inaccuracies are penalized more heavily than readability issues: 
\begin{equation}
max(R_{syn}) < min(R_{sem})\; \text{and} \; max(R_{sem}) <min(R_{read})
\label{ruleprserve}
\end{equation}
To generate $r_i$, specifically, \circled{1} \syss checks for syntax errors (See \Cref{syntaxdesign}).
If not passed, it returns a syntax penalty score.
\circled{2} Otherwise, it verifies semantic equivalence between $o_i$ and the original binary, $bin$. 
If $o_i$ fails the semantic check, \syss returns a semantic penalty score, based on the symbolic matching with the $bin$ (See \Cref{semanticdesign}).
\circled{3} Only when $o_i$ passes both checks, \syss defers to the readability component, compares $o_i$ to the original decompiled code, $o_g$, and returns a readability score (See \Cref{readabilitydesign}).
This design ensures \syss only awards readability once the accuracy has been verified, thereby adhering to our core principle and maintaining decompiled code trustworthiness.
\subsubsection{Syntax Metric}
\label{syntaxdesign}
As prior work observes~\cite{Wong2023RefiningDC,Muqi2024Dhelix}, decompiler outputs frequently are not directly recompilable even at the function level, necessitating nontrivial post-processing before they can be compiled by standard toolchains.
To address this and feed cleaner code into downstream semantic validation, \syss adopts an iterative recompilation on each function: it places each decompiled function into its own translation unit, repeatedly compiles it, and fixes the compiler-reported syntax issues.
Concretely, \syss automatically initializes undefined variables, injects required system libraries, and translates special pseudo-instructions (e.g., CONCAT) into function calls.
For each candidate code $o_i$ from the baseline model, \syss iteratively submits it to the compiler and obtains a syntax score,  $R_{syn}$, based on the compiler's feedback.
Specifically, \syss assigns $R_{syn}(o_i)$ to $syn_{pen}$ when the $o_i$ cannot be recompiled and $pass$ otherwise:
\begin{align}
R_{syn}(o_i) &= \left\{
\begin{array}{l}
syn_{pen}, \text { if } o_i \text { cannot be compiled. } \\
pass, \text { otherwise } \\
\end{array}\right.
\end{align}
Note that $syn_{pen}$ contributes to \syss, a syntax error ensures a clear penalty.

\subsubsection{Semantics Metric}
\label{semanticdesign}

Once the candidate code $o_i$ passes the syntax check, \syss runs symbolic execution to conduct the semantic check between the binary and the intermediate representation (IR) code of $o_i$.
To better describe, we split into: return value and external function call.

To confirm that the return values of the original binary and any candidate code match across different inputs, we employ symbolic execution and \texttt{SMT} solver.
Specifically, as shown on the left side of \Cref{semanticframework} in the Appendix, given the candidate code, \syss first converts each function argument into a symbolic variable and performs symbolic execution to build a \texttt{Symbolic-Model-Ret}.
This model captures the function’s semantics by linking symbolic inputs to arguments and symbolic outputs to return values, effectively encoding how each conditional transformation affects the final result.
After that, \syss runs the solver on \texttt{SMT-RET} to compare symbolic models between the original binary and the decompiled code.

When verifying external function calls, however, relying on the arguments provided from the LLM-generated candidate code (i.e., $o_i$) can yield many false negatives.
The fundamental issue is that the binary itself does not encode any definitive ground truth regarding the arguments of external function calls. 
As a result, if we model or validate such calls based solely on the argument lists produced by the LLM, the assessment becomes highly unreliable. 
In particular, an LLM can trivially circumvent these checks by generating calls with incomplete or even empty argument lists, which may still pass superficial validation despite being semantically incorrect.
More broadly, accurately recovering function prototypes for external calls remains an open and well-recognized challenge in binary analysis and decompilation~\cite{whenfuncsig2021lin}. 
The number of arguments, their types, and their semantics are typically lost during compilation and are not recoverable from the binary alone without additional assumptions or debugging information. 
Consequently, this information cannot serve as a trustworthy reference during semantic verification.

Hence, to model the external function call in LLM-generated code in a better way, \syss adopts a strategy that does not rely on unavailable prototype information. 
To do this, \syss checks whether the number of external function calls being invoked, under varying input conditions, matches between the original binary and the LLM-generated code, as shown on the right side of \Cref{semanticframework} in the Appendix.
Specifically, to identify valid external calls in each decompiled function, we first extract and save the names of all invoked external functions to a file as our ground truth. 
After that, during the symbolic execution of the LLM-generated code, we build a symbolic model called \texttt{Symbolic-Model-Call} to model the external function call behavior.
Similar to \texttt{Symbolic-Model-Ret}, this model is built through symbolic execution with symbolic inputs.
However, instead of tracking return values, it counts how many ground-truth external function calls are invoked along each execution path. 
This count is then used as a behavioral signature of function-call activity.
Finally, we run \texttt{SMT} solver on \texttt{SMT-CALL} to compare the function call-count symbolic models between the original binary and the decompiled code.
By using this approach, we successfully avoid relying on potentially unreliable function prototype information.

By sequentially integrating the return-value check and the external-call check, we define our semantics metric as follows:
\begin{equation}
R_{sem}(o_i,bin)=\left\{\begin{array}{l}
ret_{pen}, \text { if } (check_{ret}(o_i,bin)==false)\\
call_{pen}, \text { if }( check_{ret}(o_i,bin) == true,  \\
\quad \quad \quad \quad \text {and } check_{call}(o_i,bin)==false)\\ 
pass, \text { otherwise }
\end{array}\right.
\end{equation}
, where $check_{ret}(o_i,bin)$ is the boolean result of running \texttt{SMT} solver on \texttt{SMT-RET}, $check_{call}(o_i,bin)$ is the result of running \texttt{SMT} solver on \texttt{SMT-CALL}, $ret_{pen}$ and $call_{pen}$ are the penalties applied when the return‐value check or the external‐call check fails, respectively.

\subsubsection{Readability Metric}
\label{readabilitydesign}

For the reinforcement learning over multiple LLM-generated candidates, \syss must assign each candidate $o_i$ a scalar, objective readability score that supports reliable ranking.
However, existing options do not directly fit our setting.
Specifically, RLHF-style human studies primarily yield pairwise preferences (“which is better?”) rather than calibrated scalar scores, while individual software-engineering readability metrics each have limitations.
To address this limitation, we score each $o_i$ directly against the original decompiler output $o_g$ and build a hybrid assessor by customizing established software engineering readability models.
Specifically, we integrate selected features from the only decompiler-related metric,  R2I~\cite{eom2024r2i}, and the most classic readability metric, B\&W~\cite{buse1010coderead} via a weighted linear combination (i.e., multiply each feature count by its weight and sum), following the same principle used by both R2I and B\&W.
Moreover, to blend them effectively, we normalize their individual scores to a fixed interval, allowing precise control over their relative influence. 
We leverage the fact that both B\&W and R2I are defined and validated via user studies \cite{eom2024r2i,buse1010coderead}, reflecting readability features of decompiled code such as the number of references and do-while constructs.

To compute $ R_{b\&w}(o_i,o_g)$, we follow the B\&W framework,  which defines a feature set, $\textbf{f}_{b\&w} = \{f_1,f_2,...\}$, e.g., average number of commas per line, and their associated weights, $\textbf{w}_{b\&w} = \{w_1,w_2,...\}$. 
Specifically, we first apply B\&W to $o_i$ and $o_g$ to obtain two absolute scores.
We then compute their relative difference and pass this value through a sigmoid function to map it into the range [-1,1], yielding a normalized, relative readability score.
Formally,
\begin{equation}
\begin{aligned}
R_{b\&w}(o_i,o_g) = &sigmoid(\frac{R_{mul}(o_g)-R_{mul}(o_i)}{min(R_{mul}(o_i), R_{mul}(o_g))} ), \\
& \text{where} \quad R_{mul}(x) = (\sum \textbf{f}_{b\&w}(x)\cdot\textbf{w}_{b\&w})
\end{aligned}
\end{equation}

\noindent
Nevertheless, certain decompilation-specific features remain outside the scope of $R_{b\&w}(o_i,o_g)$.
Fortunately, $R_{R2I}(o_i,o_g)$ includes these features, $\textbf{f}_{R2I}$ and their associated weights $\textbf{w}_{R2I}$.
Our implementation follows the methodology established in the original paper to generate the $R_{R2I}(o_i,o_g)$.
Through renormalization of the feature-associated weights, $R_{R2I}(o_i,o_g)$ produces values within the range [-1,1], calculated using the following approach:
\begin{equation}
\begin{aligned}
R_{R2I}&(o_i,o_g) =  \sum \textbf{w}_{R2I} \cdot r_{elog}(\textbf{f}_{R2I}(o_i)-\textbf{f}_{R2I}(o_g))  ,\\
 &  \text{where} \quad r_{elog}(x) = r \cdot e^{-\log _{10}\left(1+x\right)} \\
 &  \quad \quad \quad \quad+ (1-r)\cdot \left(1-e^{-\log _{10}\left(1+x\right)}\right)
\end{aligned}
\end{equation}
, where $r$ is a binary indicator taking values in {0,1} to distinguish two types of feature groups (r=1, when features belongs to smaller values imply better readability, r=0, when larger values imply better readability).
By combining B\&W and R2I, we define an overall readability score in which $\gamma$ and $\delta$ weight each metric’s contributions as follows:
\begin{equation}
R_{read}(o_i,o_g)=\gamma \cdot R_{b\&w}(o_i,o_g)+ \delta \cdot R_{R2I}(o_i,o_g)
\label{readabilityequ}
\end{equation}

\section{Implementation}
\sys is a Python framework, using  Ghidra (v11.2).

\subsection{Reinforcement Learning}
We run TRL~\cite{trlvon} (version 0.18) from Huggingface to train models. 
We set \textit{num\_generations} to 3, \textit{num\_iterations} to 10, \textit{per\_device\_train\_batch\_size} to 3, \textit{vllm\_gpu\_memory\_utilization} to 0.7,  leaving all other parameters as default.

\subsection{Semantic Check-related Settings}
For the \texttt{Recompiler}, we set the maximum iteration to 10. For the semantics checking, we implement our semantic check framework on angr~\cite{angr} (version 9.2), z3~\cite{z3} (version 4.13), and prompt~\cite{YB2020} (version 1.0).
To reduce the training time, we set the execution timeout to 30 seconds.

\subsection{\syss Metric Settings}
We set $syn_{pen}$ to -3, $ret_{pen}$ to -2, and $call_{pen}$ to -1.5.
For readability metric, we set $\gamma$ to 0.25 and $\delta$ to 0.75 for \Cref{readabilityequ}, making $R_{read}(o_i,o_g) \in (-1,1)$.
These values follow \Cref{ruleprserve} and experimental observations to prevent sparse rewards~\cite{riedmiller2018learning}. 
Specifically, we selected ($\gamma$, $\delta$)=(0.25,0.75) after a modest grid search:
We swept $\gamma$  from 0.1 to 0.9 in steps of 0.05
(setting $\delta$ = 
1 - $\gamma$) and evaluated the aggregate objective on a held-out set of 100 samples. 
After that, we weight pair the summed (normalized) B\&W and R2I scores on each sample. 
The chosen setting produced the highest aggregate readability signal while maintaining non-sparse rewards for RL training.
\section{Evaluation}
\label{evaluationsection}
We organize this section as follows: In \Cref{experimentsetup}, we describe our experimental setup, including the evaluation of \syss, dataset details, and baseline model selection. In \Cref{OAresult} and \Cref{OIAresult}, we demonstrate how effectively \sys enhances LLM-generated decompiled code via the above setup. In \Cref{casestudy}, we provide concrete case studies illustrating how \sys refines decompiled snippets to improve both accuracy/trustworthiness and readability.

\subsection{Experiment Setup}
\label{experimentsetup}

In this section, we begin by evaluating \syss along two dimensions, applicability and precision, in \Cref{xxxscoreevaluation}. 
After that, we describe our training and evaluation dataset in \Cref{dataset}.
Finally, we detail our baseline model selections and reasons behind them in \Cref{baselineselect}.

\subsubsection{\syss Evaluation}
\label{xxxscoreevaluation}
We conduct experiments on a cluster node with an NVIDIA A100 Tensor Core GPU (80GB)~\cite{nvidia_nvidia_nodate-1}, two 32‑core AMD EPYC 7543 CPUs, and 400 GB RAM.

\smallskip
\textit{Dataset.} Our metric evaluation uses a dataset of 1,948 decompiled functions sourced from binaries in two benchmarks widely used in software security research~\cite{Muqi2024Dhelix,Zion2024ahoy,Hu2024DeGPTOD,brumley2013native,zhuo2021osprey,Dream,Dreamplusplus,chang2024tygr,TIE}: \texttt{coreutils}~\cite{coreutils} (v9.5) and \texttt{util-linux}~\cite{util-linux} (v2.41).
Specifically, we first use GCC~\cite{gcc} to compile these projects 
under x86 Linux platform and then run Ghidra~\cite{ghidra} (version 11.2) on 578 resulting binaries and object files, yielding 5,385 unique decompiled functions (1,792 from \texttt{coreutils} and 3,593 from \texttt{util‑linux}). 
To focus on functions with sufficient complexity and room for improvement, we first filter this set to include only those with at least 20 lines of code (LOC) and a cyclomatic complexity~\cite{mccabe1976complexity} greater than 3.
Nevertheless, since excessively complex or massive monolithic functions increase the binary's information entropy, which can easily be detected by heuristic indicators utilized by modern anti-virus scanners~\cite{Alshahwan2020detect}, we cap functions at less than 200 LOC.
As a result, applying these filters yields 1,948 (36.2\%) functions (653 from \texttt{coreutils} and 1,295 from \texttt{util‑linux}).

\smallskip
\textit{Methodology.} \syss contains two core components, accuracy check and readability assessment.
Because our readability metric is an aggregation of two previously validated metrics~\cite{buse1010coderead,eom2024r2i}, we accept its validity. Hence, our validity evaluation focuses on the metrics of decompiled code accuracy. 
To do this, we prompt \texttt{Qwen-Coder-2.5-3B}~\cite{Bai2023QwenTR} with the above 1,948 decompiled functions and then pass each of its generated code completions, along with the corresponding original binary, through \syss for an accuracy check.
Specifically, we evaluate the accuracy check of \syss along two dimensions: 
(1) \textit{Applicability}: The proportion of functions for which \syss can successfully derive the symbolic models from the binary. 
(2) \textit{Precision}: Among these analyzable functions, the fraction for which \syss correctly determines that the generated code is semantically equivalent to the original.

\begin{table}[t]
\scriptsize	
\parbox{.33\linewidth}{
\centering
\begin{tabular}{l|l|r}
\thickhline
\# & Categories             & Pct. \\ 
& of Errors &\\
\hline
\multirow{2}{*}{1}  & Underlying   & 43\%       \\ 
&tool errors  & \\\hline

2  & Timeout                          & 34\%       \\  \hline

\multirow{2}{*}{3}  & \syss  & 14\%       \\ 
&errors&\\
\thickhline
\end{tabular}
\caption{The percentage of decompiled functions that cannot be analyzed by \syss due to the listed errors.}
\label{applisyss}
}
\hfill
\parbox{.65\linewidth}{
\centering
\begin{tabular}{c|c|c}
\thickhline
     \begin{tabular}[c]{@{}c@{}}Categories\\ \end{tabular}                     & \begin{tabular}[c]{@{}c@{}}Originally \\ Accurate(OA)\end{tabular} & \begin{tabular}[c]{@{}c@{}}Originally \\ Inaccurate(OIA)\end{tabular} \\ \hline
(Training, & (300, &(0, \\
Evaluation)       & 725)                                                          & 548)                                                               \\ \hline
(Syntax Errs, & (0, &(307, \\ Semantic Errs) & 0)                                                              & 241)                                                             \\ \thickhline
\end{tabular}
\caption{Dataset overview. The table reports the number of functions per category, including counts used for training and evaluation, and the incidence of each error type in the original decompiled code.}
\label{tab:dataset}
}

\end{table}

\smallskip
 \textit{Applicability result.}
\syss successfully finishes the accuracy check of 80.7\% (1,573/1,948) of functions, i.e., generates the symbolic models of these functions and gets the result from the \texttt{SMT} solver without errors.
To understand the limitation, we randomly sample 50 functions, where \syss fails, without replacement, and manually analyze them. 
\Cref{applisyss} categorizes the failures encountered by \syss and reports their occurrence frequency.
These errors fall into three categories:
(1) Errors from the underlying tool, e.g., bugs in angr or incorrect identification of function boundaries when encountering no‑return calls, 
(2) Timeout due to the scale of a function, and (3) Internal errors of \syss, e.g., unsupported floating‑point instructions.

\smallskip
\textit{Precision result.} 
We evaluate the precision of \syss by randomly sampling 100 functions without replacement from the 1,573 that \syss can analyze, and manually compare each LLM-generated output against its original binary’s semantics to verify whether the decision made by \syss was correct.
\Cref{tab:scoresyss} presents our results in terms of accuracy, precision, recall, and F1 score. 
Our analysis uncovers that:
(1) False positives occur because the memory model of \syss allocates fresh memory addresses for symbolized pointers.
Even when the generated code is semantically correct, the values of pointers differ from those in the original binary, causing the symbolic executor to report a spurious mismatch.
(2) False negatives occur because we assume all external function calls return a constant value. 
When using a constant return value in a conditional instruction, one branch may never be explored during the symbolic execution of the binary. However, since users can vary the assumed constant, rerunning \syss with multiple return values may exercise most branches and thus eliminate these false negatives.

\subsubsection{Dataset}
\label{dataset}
We present a dataset overview, then explain the training and evaluation datasets.

\begin{table}[]

\scriptsize
\parbox{.25\linewidth}{

\centering
\begin{tabular}{c|c}
\thickhline
Accuracy   & 0.9450  \\ 
Precisions & 0.9100 \\ 
Recall     & 0.9785 \\ 
F1         & 0.9430  \\ 
\thickhline
\end{tabular}
\caption{ \syss on LLM output.}
\label{tab:scoresyss}

}
\hfill
\parbox{.7\linewidth}{
\centering
\begin{threeparttable}
        \begin{tabular}{c|l|c|c|c}
            \thickhline
            \# & Model Name & Synt Errs & Sem Errs & Total Errs \\
            \hline
            \multirow{2}{*}{1} & Qwen2.5-   & 288  & 61  & 349 \\
             & Coder-1.5B   &  (39.7\%)  &  (8.41\%) &  (48.1\%)\\
            \multirow{2}{*}{2} & Qwen2.5-     & 196  & 83 & 279  \\
               & Coder-3B     &  (27.0\%) &  (11.4\%)&  (38.5\%) \\
            \multirow{2}{*}{3} & Llama3.2         & 279   & 114 & 393 \\
             & -3B          &  (38.5\%)  &  (15.7\%)&  (54.2\%)\\
            \multirow{2}{*}{4} & LLM4Dec-    & 636  & 46  & 682 \\
              & ompile-1.3B        & (87.7\%) &  (6.34\%) &  (94.0\%)\\
            \thickhline
        \end{tabular}
    \end{threeparttable}
    \vspace{3pt}
\caption{Different baseline models' performance on the \texttt{(OA)} dataset.}
\label{tab:baselineresult}
}

\end{table}

\smallskip
\textit{Dataset overview.} Given that \syss supports 1,573 functions (see \Cref{xxxscoreevaluation}), we center our overview on this filtered set.
To better understand our dataset, we run \syss on the raw Ghidra outputs of these functions and examine their results distribution.
Based on the accuracy check, the functions were split into two groups: originally accurate (i.e., error-free) and originally inaccurate. 
\Cref{tab:dataset} shows the results: 
Of the 1,573 functions, 1,025 are classified as originally accurate (i.e., pass \syss's accuracy check), while the remaining 548 fall into originally inaccurate (i.e., fail the \syss's accuracy check), 307 exhibiting syntax errors and 241 containing semantic errors.

\smallskip
\textit{Training and evaluation.}
We randomly choose 300 functions from the originally accurate group as the training data. 
This selection criterion is based on the fact that LLMs demonstrate a limited ability to correct decompiler-introduced inaccuracies when provided only with decompiled code, as listed in \Cref{OIAresult}.
We apply this template uniformly during both training and inference: \textit{``prompt'': [
            \{
            ``role'': ``system'',
                ``content'': ``You are a helpful assistant for improving the decompiled result from the user. The user will input the decompiled result from Ghidra. Please improve its readability while preserving its semantics. Please do not add comments. Please just output the improved code.''
            \},
            \{``role'': ``user'',
                ``content'': ``[original decompiled code]''\}}.

We evaluate \sys on the remaining 1,273 functions with two subsets: 725 functions that are \textbf{originally accurate (OA)} and 548 functions that are \textbf{originally inaccurate (OIA)}.

\subsubsection{Baseline Model}
\label{baselineselect}
Due to computational resource constraints, we restrict our LLM to fewer than 3 billion parameters.
Specifically, as a reinforcement learning framework, GRPO is particularly memory-intensive, since it requires simultaneous inference to generate candidate outputs and backpropagation to update model parameters, both of which consume significant GPU memory.  
Additionally, the reward normalization in GRPO mandates generating at least two candidates per input, further increasing memory demands. 
In our experiments, even after using vLLM~\cite{kwon2023efficientmemorymanagementlarge} with a memory‐efficient setting of 0.7, we still encountered out‐of‐memory errors when running on models larger than 3 billion parameters.

Therefore, we select two code-focused LLMs, \texttt{Qwen2.5-Coder}~\cite{Hui2024Qwen25CoderTR} (\texttt{1.5B} and \texttt{3B} variants) and one general-purpose model, \texttt{Llama3.2-3B}~\cite{grattafiori2024llama3herdmodels}.
As shown in \Cref{tab:baselineresult}, due to the poor performance of \texttt{LLM4Decompile-End-1.3B}~\cite{Tan2024LLM4DecompileDB} on our dataset, where 93\% of the functions go wrong after applying it, we excluded it from our main evaluation.
More specifically, we ask all four above models to improve the functions in our \texttt{OA} evaluation dataset, score the outputs using \syss 
and show the results in \Cref{tab:baselineresult}. 
Since being fine-tuned based on the model released from November 2023, \texttt{LLM4Decompile-End-1.3B} made 682 functions (94.0\%) inaccurate.
In contrast, the three newer LLMs (released around September 2024) 
made only 46.9\% of functions inaccurate.
Given this marked difference, we omit \texttt{LLM4Decompile-End-1.3B} from further analysis.

\subsection{\sys Performance on OA}
\label{OAresult}

We show the evaluation of \sys on originally accurate functions (\texttt{OA}) in two steps.
We demonstrate changes brought from \sys in \Cref{OAresultsub}. 
\Cref{OAfind} analyzes its underlying reasons.

\begin{figure}[h]
\begin{subfigure}{0.85\columnwidth}{
  \centering
  \includegraphics[width=1\linewidth]{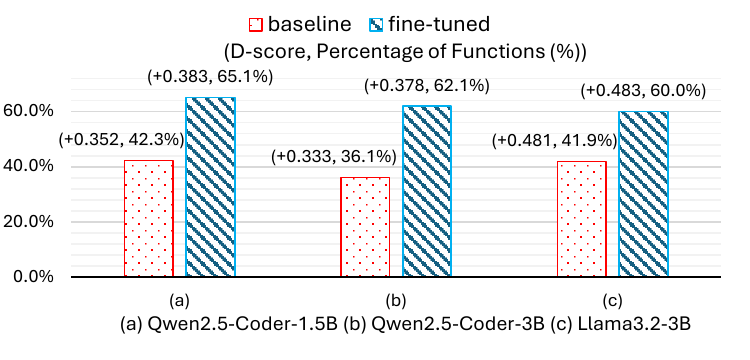}
  \caption{The performance of different models on the \texttt{OA} dataset, before and after fine-tuning by \sys.}
   \label{tab:oaresult}
}\end{subfigure}
  \hfill
\begin{subfigure}{0.85\columnwidth}{
  \centering
{\includegraphics[width=1\linewidth]{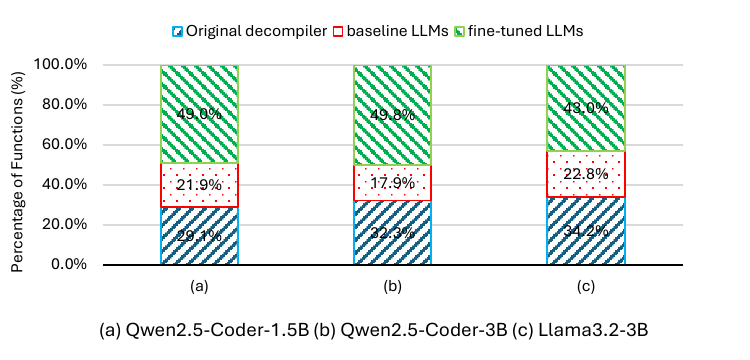}}
\caption{Overall performance of the selection system within \sys on the \texttt{OA} dataset.}
\label{fig:3in1compare}
}\end{subfigure}
\end{figure}

\subsubsection{Result} 
\label{OAresultsub}
To clearly present, we split the result into two parts: (1) baseline and fine-tuned LLMs' comparisons, and (2) an overall performance assessment.

\textit{Fine-tuned LLMs vs. baseline LLMs.} \Cref{tab:oaresult} summarizes the number of original decompiled functions within the \texttt{OA} dataset that are improved by the baseline and fine-tuned models and their average \syss.
To this end, we compare the output from each model with the original decompiled code.
Compared with the baseline LLMs without \syss-driven fine-tuning, our trained LLMs on average improve 55.3\% more decompiled functions. 
By selecting the best output among all three baseline LLMs, 509 (70.2\%) functions (avg. +0.465) show improvement, while fine-tuned LLMs achieve improvement in 616 functions (84.9\%), averaging +0.517.

\textit{Overall improvement by \sys.}
\Cref{fig:3in1compare} shows the result of the selection system (i.e., selecting the version with the highest \syss score among the fine-tuned model, the baseline model, and the original decompiled output).
Specifically, for \texttt{Qwen2.5-Coder-1.5B}, fine-tuned LLM's output is the best for 355 (49.0 \%) functions (avg. +0.399), baseline LLM for 159 (21.9 \%) functions (avg. +0.419).
For \texttt{Qwen2.5-Coder-3B}, fine-tuned LLM's output is best for 361 (49.8\%) functions (avg. +0.402) and baseline LLM for 130 (17.9\%) functions (avg. +0.398).
For \texttt{Llama3.2-3B}, fine-tuned LLM's output is best for 312 (43.0\%) functions (avg. +0.523) and baseline LLM for 165 (22.8\%) functions (avg. +0.561).
Moreover, by taking the best output among all six LLMs and the original decompiled code, 625 (86.2\%) functions showed improvement (avg. +0.585), while 100 functions remained unimproved.

\begin{table}[H]
\centering
\tiny
\begin{tabular}{c|c|rrrr|rrrr}
\thickhline
\multirow{2}{*}{\#} & \multirow{2}{*}{Model Name} & \multicolumn{4}{c|}{Improvements}                                                                      & \multicolumn{4}{c}{Regressions}                                                                       \\ \cline{3-10}  
                    &                             & \multicolumn{1}{c}{Syn} & \multicolumn{1}{r}{Sem} & \multicolumn{1}{c}{Read} & Tot & \multicolumn{1}{c}{Syn} & \multicolumn{1}{c}{Sem} & \multicolumn{1}{c}{Read} & Tot \\ \hline
1                   & Qwen2.5-Coder-1.5B          & \multicolumn{1}{r|}{135}    & \multicolumn{1}{r|}{18}       & \multicolumn{1}{r|}{202}         & 355   & \multicolumn{1}{r|}{25}     & \multicolumn{1}{r|}{2}        & \multicolumn{1}{r|}{132}         & 159   \\ 
2                   & Qwen2.5-Coder-3B            & \multicolumn{1}{r|}{88}     & \multicolumn{1}{r|}{17}       & \multicolumn{1}{r|}{256}         & 361   & \multicolumn{1}{r|}{12}     & \multicolumn{1}{r|}{2}        & \multicolumn{1}{r|}{116}         & 130   \\ 
3                   & Llama3.2-3B                 & \multicolumn{1}{r|}{116}    & \multicolumn{1}{r|}{38}       & \multicolumn{1}{r|}{114}         & 268   & \multicolumn{1}{r|}{19}     & \multicolumn{1}{r|}{9}        & \multicolumn{1}{r|}{150}         & 178   \\ \thickhline
\end{tabular}
\caption{This table shows how many functions get improved and regressed by the training of \sys.}
\label{tab:improvenum}
\end{table}

\subsubsection{Findings}
\label{OAfind}
To better understand the training part of the \sys's influence on LLM performance, we conducted an additional analysis by comparing the fine-tuned output and the baseline model output.
Specifically, we analyze how the training within \sys modifies function performance through two distinct categories: improvements, where fine-tuned models successfully resolve issues present in baseline outputs, and regressions, where previously error-free baseline functions develop new problems following the fine-tuning process.
Based on it, we classify these changes into six categories: syntax fixes, semantic fixes, syntax regressions, semantic regressions, readability improvements, and readability regressions.
\Cref{tab:improvenum} reports the number of functions falling into each category. 
These results show that LLMs fine-tuned by \sys achieve considerable accuracy enhancements, with regression instances remaining at acceptably low levels.

To investigate the root causes of both improvements and regressions, we randomly selected 50 functions from each group (i.e., those that improved and those that regressed) and performed a manual code review. 
Our findings show that fine-tuning corrects five main categories of errors (\Cref{tab:rootimprove}): 
(1) Missing instructions, e.g., missing goto labels, variable declarations, and value assignments.  
(2) Incorrect brackets in conditional expressions, e.g.,  missing parentheses in if statements or misplacement leading to incorrect pointer dereference. 
(3) Incorrect variable naming and casting, e.g., using \textit{mode\_t} when only \textit{\_\_mode\_t} is defined.
(4) Unwanted instruction insertions, e.g., duplicate goto labels and unwanted function calls.
(5) Incorrect literal value. e.g., constant values are inconsistent with the original values.
However, fine-tuning also introduces regressions in three areas: 
(1) Missing instructions.
(2) Incorrect brackets in conditional expressions.
(3) Unwanted instruction insertions.
More examples of the enhancements that \sys enables are in \Cref{casestudy}.

\begin{center}
    \fbox{
    \parbox{0.93\columnwidth}{\textbf{Summary}: When the original decompiled code does not have syntax or semantic errors, \sys yields substantial improvement, in both the number of functions improved and their average \syss.}
    }
\end{center}

\begin{table}[t]
\tiny
\parbox{.37\linewidth}{
\centering
\begin{tabular}{c|c|r}
\thickhline
Perform                  & Error Type                            & No. \\ 
\hline
\multirow{5}{*}{Improve} & Missing instruction                   & 24          \\  
                             & Incorrect brackets                    & 12          \\ 
                             & Incorrect casting & 5           \\ 
                             & Unwanted instruction       & 7           \\  
                             & Incorrect literal value               & 2           \\ \hline
\multirow{3}{*}{Degrade} & Missing instruction                   & 14          \\ 
                             & Incorrect brackets                    & 24          \\ 
                             & Unwanted instruction        & 12          \\ 
\thickhline
\end{tabular}

\caption{The root causes of both improvements and regressions caused by \sys.}

\label{tab:rootimprove}
}
\hfill
\parbox{.58\linewidth}{
\centering
 \begin{threeparttable}
        \begin{tabular}{c|c|c|c}
            \thickhline
            \# & Model Name & No. of Func & {Avg. D-S.} \\
            \hline
            \multirow{2}{*}{1} & Qwen2.5-Coder   & 8 & \multirow{2}{*}{-0.591} \\
             & -1.5B-baseline   & (1.48\%)  & \\
             
            \multirow{2}{*}{2} & Qwen2.5-Coder     & 12 &  \multirow{2}{*}{-0.936} \\
             & -3B-baseline     &  (2.23\%) &  \\
            
            \multirow{2}{*}{3} & Llama3.2         & 14 (2.6\%)  &  \multirow{2}{*}{-0.614} \\
              & 3B-baseline          &  (2.6\%)  &  \\
             
            \multirow{2}{*}{4} & All base-           & 31  & \multirow{2}{*}{-1.170} \\
             & line models           &  (5.66\%) & \\
            
            \multirow{2}{*}{5} & Qwen2.5-Coder & 6   & \multirow{2}{*}{+0.113} \\
             & -1.5B-fine-tuned &  (1.12\%)  & \\
            
            \multirow{2}{*}{6} & Qwen2.5-Coder   & 9  &  \multirow{2}{*}{-0.396} \\
             & -3B-fine-tuned   & (1.67\%)  &  \\
            
            \multirow{2}{*}{7} & Llama3.2        & 8   & \multirow{2}{*}{-0.243} \\
             & -3B-fine-tuned        & (1.49\%)  &  \\
            
            \multirow{2}{*}{8} & All fine-         & 19  & \multirow{2}{*}{-0.469} \\
              & tuned models         & (3.47\%) &  \\
            \thickhline
        \end{tabular}
    \end{threeparttable}
    
    \vspace{3pt}
\caption{Different models' performance on the \texttt{OIA} dataset, before and after fine-tuning by \sys.}
\label{tab:oiaresult}

}
\end{table}
\subsection{\sys Performance on OIA}
\label{OIAresult}

We show the evaluation of \sys on originally inaccurate functions (\texttt{OIA}) in two steps. We first compare outputs from the baseline and fine-tuned models and then analyze the underlying reasons for these observed scores.

\subsubsection{Result}
\label{OIAresultsub}
\Cref{tab:oiaresult} summarizes, for both baseline and fine-tuned models, how many functions achieve a higher \syss than the original decompiled output and the average score improvement for those functions. As shown, both the baseline model and the fine-tuned models struggle with improving the original inaccurate functions.
Moreover, by selecting the best among all six LLMs' output and the original decompiled code, only 44 functions showed improvement (average -0.759), while 504 functions remained unimproved (average -2.50).

\subsubsection{Findings}
\label{OIAfind}
To investigate why our LLMs perform poorly on the OIA dataset, we randomly select 50 functions, where 25 are successfully improved by any LLMs (including baseline LLMs and fine-tuned LLMs) and 25 are not improved by LLMs, and manually analyze the underlying factors.

For the 25 functions where LLMs achieve improvements, surprisingly, we observed a consistent pattern: each involved variables declared as \texttt{undefined [16]}, a 16-byte type with unknown signedness. 
The LLMs converted the variables with these opaque declarations into fixed-length arrays and updated member accesses.
For instance, we observe that the declaration changed from \texttt{undefined [16] auVar1;} to an array, \texttt{ulong auVar1[2]} and this variable's first eight-byte assignment is modified from \texttt{auVar1.\_0\_8\_ = 0; } to \texttt{auVar1[0] = 0;}.
Although these transformations sometimes introduce subtle semantic inaccuracies, they consistently eliminate syntax errors, resulting in higher \syss.

For the 25 functions where LLMs fail to achieve any improvements, we find three main causes: (1) Syntax errors due to undefined function pointer types. In 13 cases, errors stem from unresolved function pointer types like \texttt{(code *)puVar3[4]}. (2) Syntax errors from unsolvable type patterns. In 7 cases, issues arise from constructs like \texttt{CONCAT31((int3)XX)}, where the type \texttt{int3} is not well-formed. (3) Semantic errors from uninitialized global variables. In 5 cases, the decompiled code fails to initialize global variables to the correct value.

\begin{center}
    \fbox{
    \parbox{0.93\columnwidth}{\textbf{Summary}: When the original decompiled code contains syntax or semantic errors, neither baseline LLMs nor our fine-tuned LLMs can deliver real improvements, indicating the importance of the decompiler front-end and the limitation of current LLMs in fixing those errors introduced by the decompiler.}
    }
\end{center}


\subsection{Case Study}
\label{casestudy}

In this section, we present several illustrative cases that demonstrate how \sys effectively improves the readability of decompiled code while preserving the accuracy.
Specifically, we show how \sys fixes syntax errors introduced by the LLM in \Cref{syntaxcase}. 
Moreover, we provide an example about how \sys fixes semantic errors introduced by the LLM in \Cref{semanticcase} in the Appendix.


\subsubsection{Syntax Error Fixes}
\label{syntaxcase}
\noindent\textbf{\textit{Fixing Incorrect Variable Casting and Refactor Loop.}} \Cref{fig:casestudy3} in the Appendix illustrates how \sys enhances the output of \\\texttt{Qwen2.5-Coder-3B} model for the decompiled snippet of the \texttt{uuid\_copy} function (\texttt{util-linux}).

Specifically, this function simply copies 16 bytes from the second parameter (\texttt{src}) to the first (\texttt{dst}). 
Since the decompiler cannot reconstruct the original struct or pointer types for the two input parameters, it falls back to manual pointer casting these parameters to byte pointers and uses pointer offsetting by (\texttt{lVar1}) to copy each byte (see line 8 of the original decompiled output). 

Meanwhile, the decompiler also uses a \texttt{do-while} loop that checks \texttt{lVar1 != 0x10}, rather than the more natural \texttt{for (i = 0; i < 16; i++)} construct.
Note that implementing a rule or heuristic inside Ghidra to systematically convert such \texttt{do-while} loops into \texttt{for} loops is non-trivial. In particular, Ghidra first normalizes \texttt{for} loops into \texttt{while} loops in \texttt{ActionStructureTransform}, and in this case the loop is not even recognized as a \texttt{while} loop because it fails the checks in \texttt{CollapseStructure::ruleBlockWhileDo}, which relies on several heuristics (e.g., the condition block must have exactly two outgoing edges, and a body-start block should typically have only one incoming/parent block). 

In contrast, an LLM can easily convert this pattern into a canonical \texttt{for} loop, and the correctness of such a transformation can be validated by \syss.
Specifically, when the baseline LLM refactors the loop, though it successfully rewrites it as a \texttt{for} with \texttt{i < 16}, it mistakenly applies the array subscript operator to the input parameters of type \texttt{long}, leading to the compiler error.
\sys, however, fine-tunes the model to insert the correct byte-pointer casts and maintain the \texttt{for} with \texttt{i < 16}, which eliminates the syntax error while yielding a more readable, semantically faithful implementation.
As a result, \sys improves the \syss from -3.000 to +0.8887.

\textbf{\textit{Fixing Missing Parentheses and Consolidating Conditionals.}} \Cref{fig:casestudy2} in the Appendix demonstrates another example about how \sys corrects the syntax errors while enhancing readability in the \texttt{fputs\_color\_cell\_close} (\texttt{util-linux}).

Specifically, for inaccuracy, line 5 in baseline \texttt{Qwen2.5-Coder-1.5B} model output lacks the necessary parentheses, resulting in the compilation error, \textit{lvalue required as left operand of assignment}.
Regarding readability, the original source code combines three checks with two \texttt{or} expressions (line 1).
The original decompiled output, however, expands this into three separate \texttt{if} statements. 
For the output from the baseline model, though it removes redundant braces, it still uses three \texttt{if} blocks.
Our fine-tuned model not only adds the necessary parentheses in line 4 but also merges the second and third checks into one consolidated if statement in line 2, mirroring the source code’s succinct logic. 
As a result, \sys raises the \syss for this function from –3.000 to +1.275.~\looseness=-1

\section{Discussion}
\label{discussion}

\textit{Applicability on \texttt{OIA} dataset.}
As illustrated in \Cref{OIAfind}, \sys performs poorly when the decompiler front-end fails to produce accurate code. 
Moreover, because \sys relies on function‐level decompiled output, it cannot help if the decompiler is unable to process the binary. 
Hence, \sys performs poorly in addressing internal decompiler issues, such as bugs that cause decompiler crashes or persistent decompilation challenges, such as function boundaries identification, type recovery, or indirect calls. 
We point out that decompilation is {\em only one of multiple aspects of binary reverse engineering}, focusing on presenting users with more readable source code of the subject binary. 
The other aspects of reverse engineering are better supported by other specialized reverse engineering tools~\cite{Valgrind2007Nethercote,drmemory2011Bruening,salwan2015triton,Lin2010AutomaticRE,qibin2022augmenting,Zion2024ahoy,tianrou2024indirectcall} to complement decompilation.
Moreover, as future work, we will explore the {\em integration} of such specialized tools (some implemented as agents) into the decompilation pipeline, to iteratively improve the decompiled code in multiple aspects (e.g., types, indirect calls, etc.).


\textit{Underlying tools.}
Since \sys employs angr~\cite{angr}, klee~\cite{klee} to run the symbolic execution, some features, including floating-point instructions, double pointers (due to the memory model), and large binaries or long context decompiled code because of timeouts, are currently not supported. We leave them as future work.

\textit{Readability metrics.}
We acknowledge that current approaches for assessing code readability are not perfect. However, R2I remains the state-of-the-art available metric for automatically evaluating the quality of decompiled code. This automation is a critical prerequisite for our RL pipeline. Furthermore, our framework is designed to be extensible; it can seamlessly integrate superior metrics as they emerge in the future. Since the development of novel readability metrics falls outside the scope of this paper, we leave it for future work.

\section{Related Work}
Regarding LLM-based decompiled-code enhancement, several approaches have been proposed.
Besides LLM4Decompile~\cite{Tan2024LLM4DecompileDB}, researchers have also proposed DecGPT~\cite{Wong2023RefiningDC}, focusing on making decompiled code more recompilable.
Nevertheless, this approach does not handle the semantic errors, e.g., hallucination errors, introduced by the LLM. 
To the best of our knowledge, DeGPT~\cite{Hu2024DeGPTOD} is the only existing work that attempts to validate the accuracy of LLM-generated decompiled code.
DeGPT introduces MSSC, a static analysis framework that assigns random values to inputs and observes the resulting changes in symbolic values in both the decompiled code and the LLM-generated code, aiming to detect discrepancies and identify inaccuracies.
However, this method has notable drawbacks. 
For instance, MSSC does not recompile the code; it cannot detect syntax errors. 
Meanwhile, by testing with random inputs, the validation result can be inconsistent across runs.

Fine-tuning LLM to generate better quality code is not new.
PPOcoder~\cite{shojaee2023execution} uses structural differences, measured via Data Flow Graphs (DFGs) and Abstract Syntax Trees (ASTs) between the source code and generated code, as its reward signal, to improve the performance of LLM in multiple code generation tasks.
StepCoder, CodeRL, and Palit~\cite{dou2024stepcoder,palit2024generating,le2022coderl} utilize compiler and unit tests as feedback for reinforcement learning.
Nevertheless, all the above approaches accept only one unique ground truth, which makes them inappropriate in the decompilation scenario.

Researchers have also investigated the use of symbolic execution tools to validate the semantics of LLM-generated code. 
Taneja~\cite{taneja2024llm} applies Alive2~\cite{alive22021lopes} on the vectorized code generated by LLM to verify its semantic correctness.
Similarly, Wang~\cite{wang2024enhancingtranslationvalidationcompiler} integrates Alive2 into a compiler’s translation-validation pipeline to ensure semantic fidelity.
Nevertheless, since Alive2 is primarily designed to detect bugs, such as undefined behaviors, arising from compiler optimizations, it may miss decompiler-specific errors.

Regarding the code readability metric, subsequent studies are proposed based on the B\&W framework. 
Specifically, researchers~\cite{holst2021importanceshortcomingscodereadability,scalab2018codereadability,posnet2011softwarread,marvin2020emperread,trockman2018utomaticalread} have shifted toward the broader concept of code comprehensibility, which extends readability by also considering elements beyond the code itself, such as associated documentation, during evaluation. 
However, because these metrics assume the presence of comments and external documentation, features that decompiled code typically lacks, they are not well suited for assessing decompiled output.

\section{Conclusion}
We design \sys, an enhanced decompiler-LLM pipeline with a fine-tuned LLM using code quality-aware RL to improve the trustworthiness and quality of the decompiled code, adhering to the principle of {\em preserving accuracy while improving readability}.
We propose \syss, an integrated scoring mechanism designed specifically for decompilation recovery tasks.
We implement \sys based on Ghidra and fine-tune three LLMs, and achieve significant decompiled code improvement for widely used benchmark functions.
Compared to the baseline LLMs, on average, our fine-tuned LLMs improve the quality of 55.3\% more functions. 
Overall, \sys generates 68.2\% better quality functions than the native decompiler, where 47.3\% of functions come from the \syss-driven fine-tuned model and only 20.9\% from the baseline model.
\section*{Ethics Considerations}

We note that our found bugs do not pose an immediate threat to users or developers since they mainly affect the accuracy of LLM-generated code.

We acknowledge the use of Chat-GPT~\cite{openai2025chatgpt} and Claude-AI~\cite{anthropic2025claude} solely as a paraphrasing aid to improve clarity and readability; at no point did we allow it to generate new content, ideas, or arguments. 
All substantive work and original insights remain entirely our own.

\clearpage
\appendix
\section{Appendix}

In this section, we show an example about how the LLM may make mistakes (\Cref{codeexample}), examine training-size
effects (\Cref{trainefficient}), compare two models with identical
architecture but different parameter counts (1.5B and 3B) (\Cref{modelcompare}), and an example about how \sys fixes the semantic error introduced by LLMs(\Cref{semanticcase}).
\begin{figure*}[!th]

        \centering
  \includegraphics[width=0.9\linewidth]{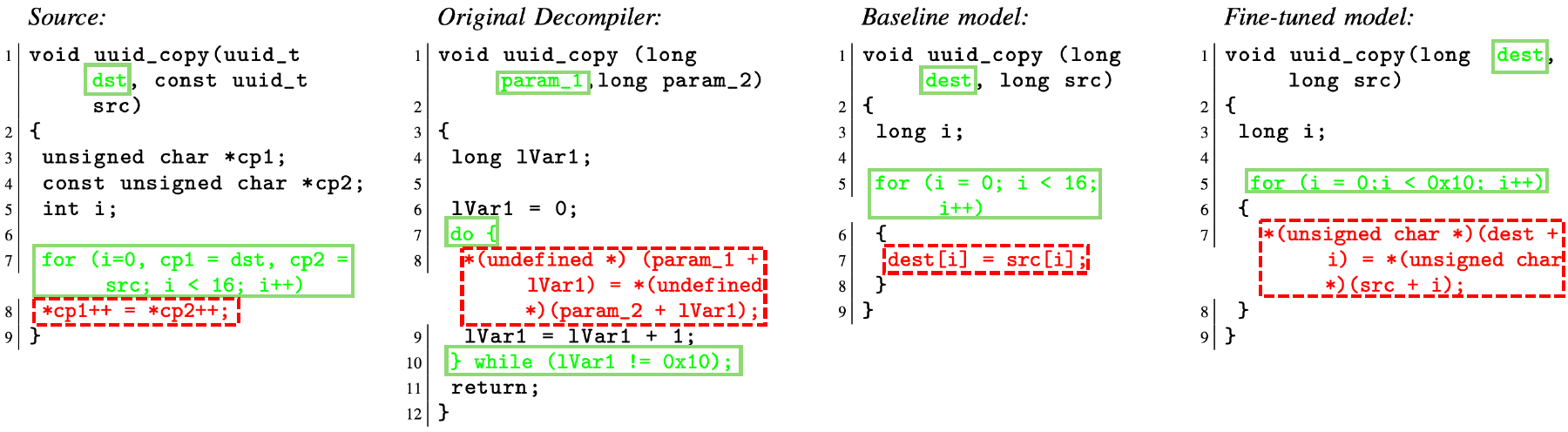}

\caption{This code snippet shows how \sys helps \texttt{Qwen2.5-Coder-3B} enhance its ability to generate more readable code (boxes with solid lines) while correcting the syntax error(boxes with dashed lines).}
    \label{fig:casestudy3}
\vspace{-0.75em}
\end{figure*}
\begin{figure*}[!th]
\centering
  \includegraphics[width=0.9\linewidth]{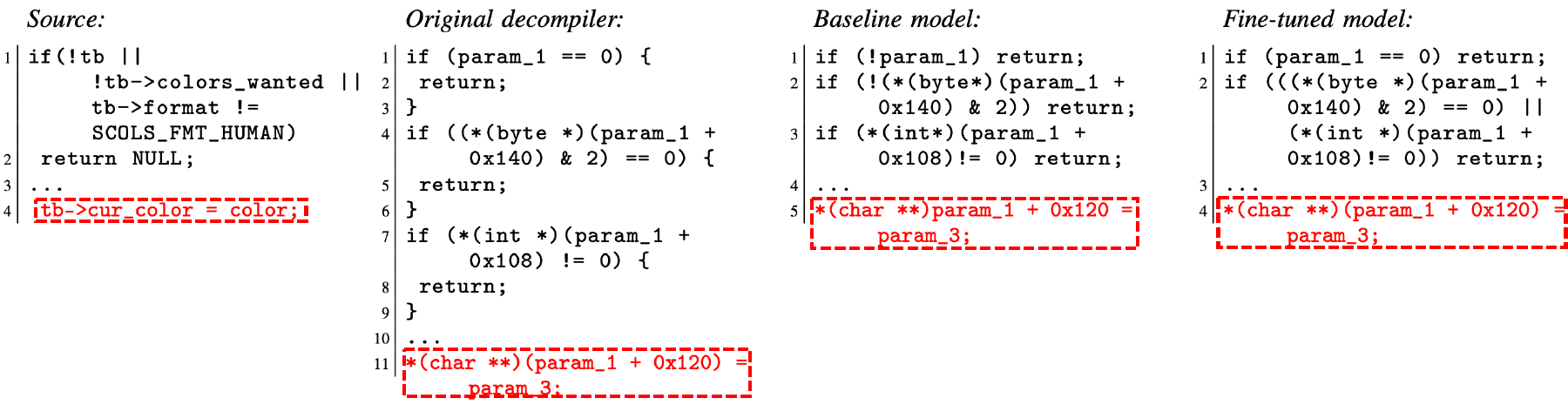}
\caption{This code snippet shows how \sys helps \texttt{Qwen2.5-Coder-1.5B} enhance its ability to generate more readable code while correcting the syntax error (boxes with dashed lines).}
    \label{fig:casestudy2}
\vspace{-0.75em}
\end{figure*}

\subsection{Code Example}
\label{codeexample}
\begin{figure*}[!th]
\label{fig:intro_code}
\centering
  \includegraphics[width=0.9\linewidth]{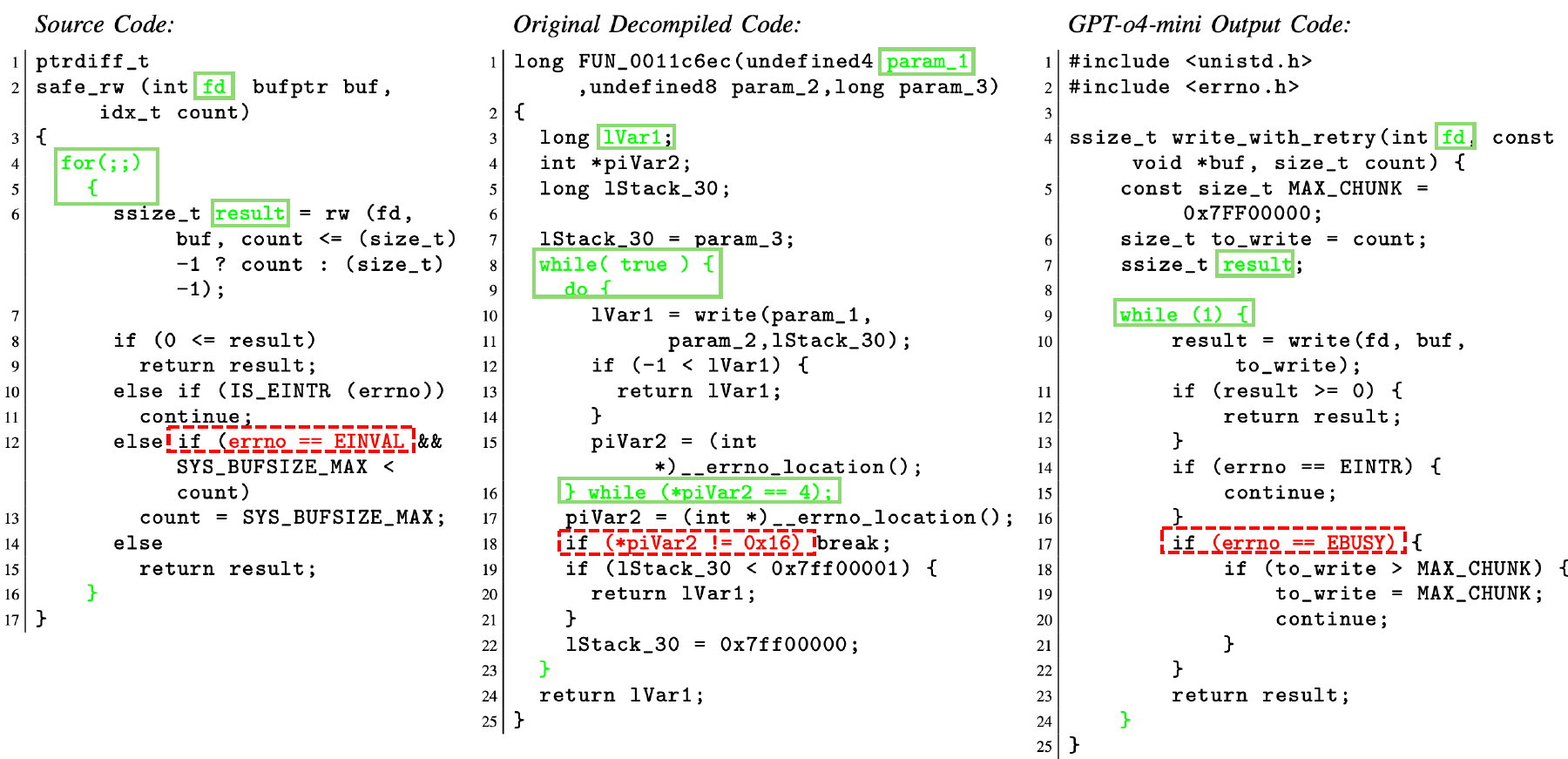}

\caption{
The code snippet shows the original decompiled code, its corresponding source code, and the version output from the GPT-o4-mini, where line 17 of the GPT-o4-mini‑refined code is misinterpreted.}
    \label{fig:introcode}
\vspace{-0.75em}
\end{figure*}
As shown in ~\Cref{fig:introcode}, while GPT-o4-mini~\cite{openai2024gpt4ocard} improves readability by recovering better variable names and collapsing nested loops into a single loop (boxes with solid lines), it erroneously replaces the condition hex value 0x16 with EBUSY instead of the correct constant (boxes with dashed lines), introducing a semantic error.

\subsection{Training Efficiency}
\label{trainefficient}

\begin{figure}{
  \includegraphics[width=0.9\linewidth]{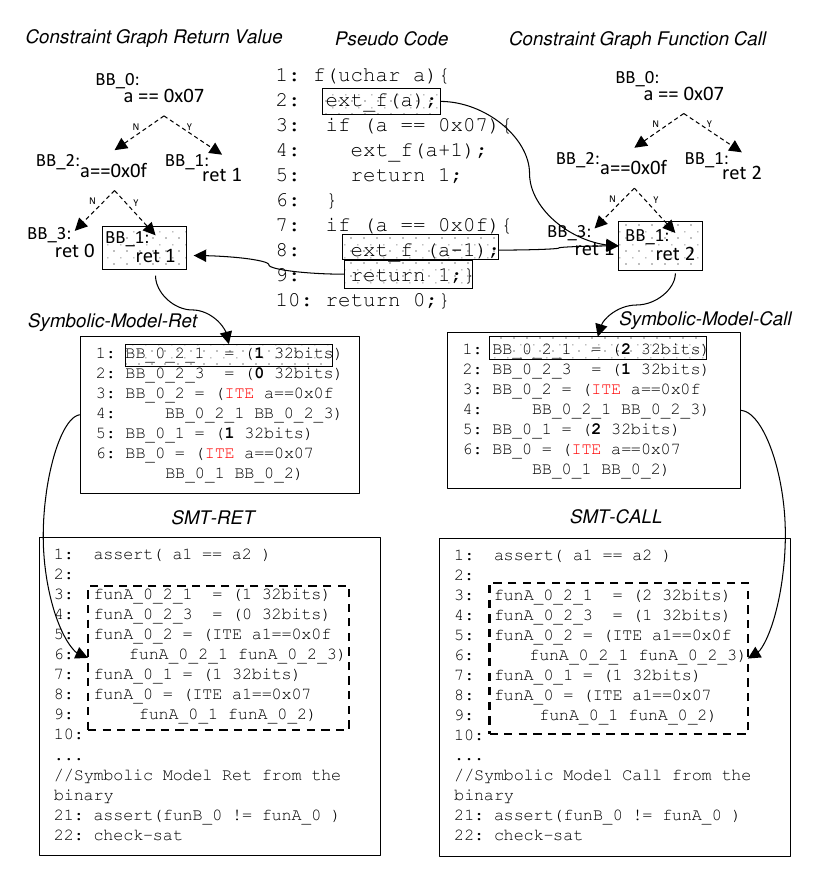}  
  \caption{The workflow of semantic check in \syss.  }
  \label{semanticframework}
}\end{figure}

\begin{figure}[htbp]
    \centering
    \begin{subfigure}[b]{0.48\textwidth}
        \centering
        \includegraphics[width=\linewidth]{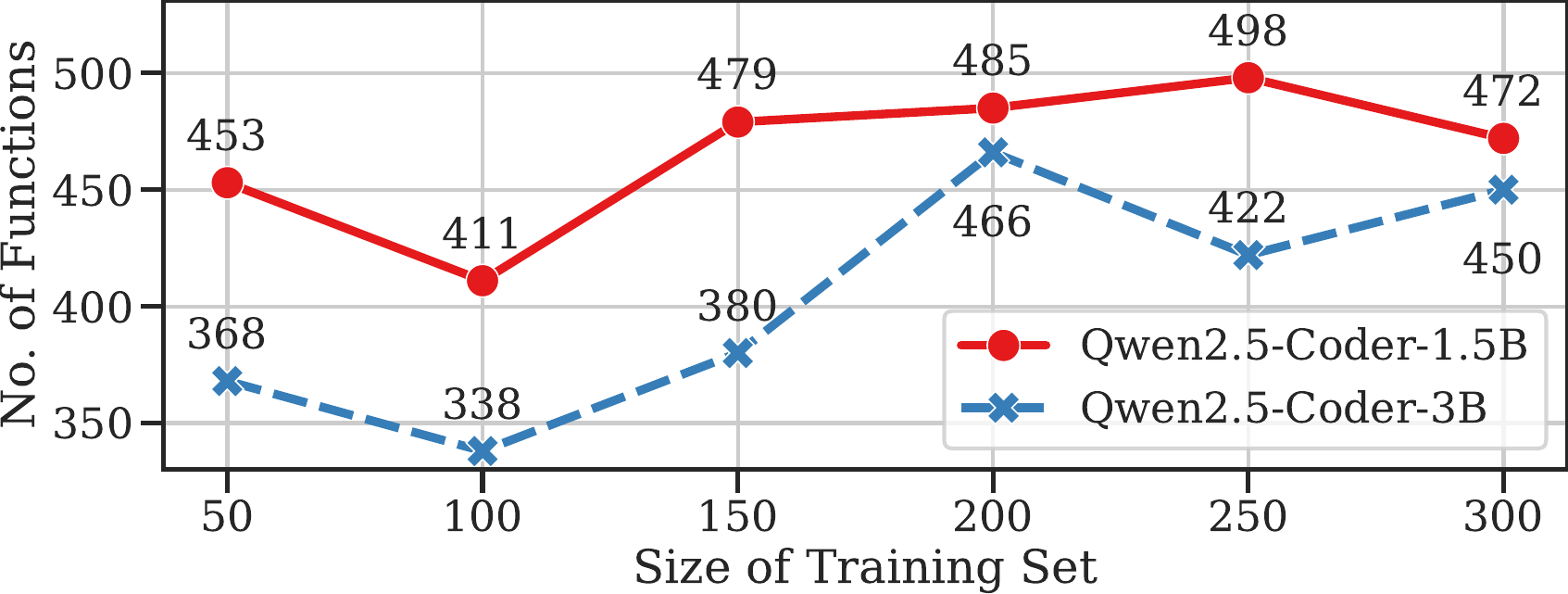}
    \caption{This table shows how many functions from the \texttt{OA} dataset scored higher (x-axis) after being processed by the fine-tuned LLMs with different sizes of the training sets (y-axis), compared to the raw decompiled code under \syss.}
   \label{tab:training}
    \end{subfigure}
    \hfill
    \begin{subfigure}[b]{0.48\textwidth}
        \centering
\includegraphics[width=\linewidth]{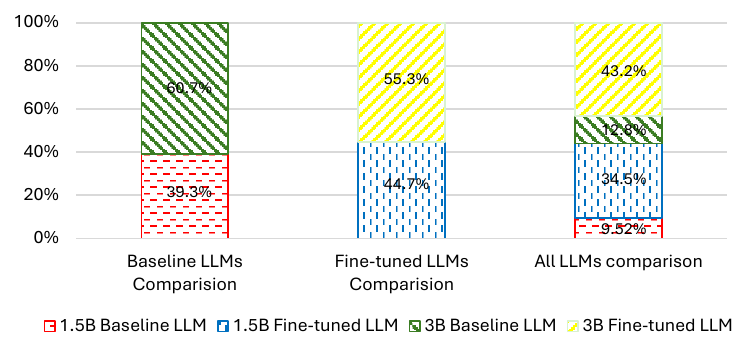}
\caption{Overall performance of the model with different size on the \texttt{OA} dataset. This chart shows, for each function, which LLM achieves the highest \syss score and reports the percentage of functions in each category.}
\label{fig:modelcompare}
    \end{subfigure}

\end{figure}

To determine the optimal training set size, we evaluated LLM performance with training sets of 50, 100, 150, 200, 250, and 300 functions.
~\Cref{tab:training} shows how many functions from the \texttt{OA} dataset scored higher after being processed by the fine-tuned LLM, compared to the raw decompiled code under \syss. 
As shown in ~\Cref{tab:training}, performance levels off once the training set reaches 200 functions.

\subsection{Model Comparison}
\label{modelcompare}

Recent studies and public benchmarks~\cite{xu2025cruxeval,Kaplan2020ScalingLF,chou2025autocodebenchlargelanguagemodels} consistently show a strong correlation between model scale and code-generation quality: larger LLMs generally score substantially higher than small models across code benchmarks, in line with established scaling laws for language modeling.
Nevertheless, our computational resource constrains our evaluation to relatively small models (Qwen2.5-Coder-1.5B and Qwen2.5-Coder-3B). 
To make the most informative comparison within this constraint, we provide a comprehensive comparison between 1.5B and 3B models under \syss. 
Specifically, on the \texttt{OA} dataset, we conduct three comparisons: (i) for the two baseline models, we select the better output for each instance and report the proportion of wins per model; (ii) we repeat the same procedure for the two fine-tuned models; and (iii) we perform an oracle-style comparison over all four models (two baselines and two fine-tuned) by selecting the best output per instance and analyzing the contribution of each model.
\Cref{fig:modelcompare} shows the result of the above three comparisons. Specifically, for baseline models, the 3B model's output is the best for 440 (60.7 \%) functions (avg. -0.998), the 1.5B model for 285 (39.3 \%) functions (avg. +0.217). 
For fine-tuned models, the 3B model's output is the best for 401 (55.3 \%) functions (avg. -0.324), the 1.5B model for 324 (44.7 \%) functions (avg. +0.340). 
Across all four models, the 3B fine-tuned model's output is the best for 313 (43.2 \%) functions (avg. -0.294), the 3B baseline model for 93 (12.8 \%) functions (avg. +0.258), the 1.5B fine-tuned model's for 250 (34.5 \%) functions (avg. -0.334) and the 1.5B baseline model for 69 (9.52 \%) functions (avg. +0.551),
Overall, the 3B models dominate the 1.5B models in both baseline and fine-tuned settings, and \sys does not alter this dominance trend, consistent with established scaling-law observations that larger models tend to achieve stronger performance.

\subsection{Semantic Error Fixes}
\begin{figure*}[th!]

\centering
  \includegraphics[width=0.98\linewidth]{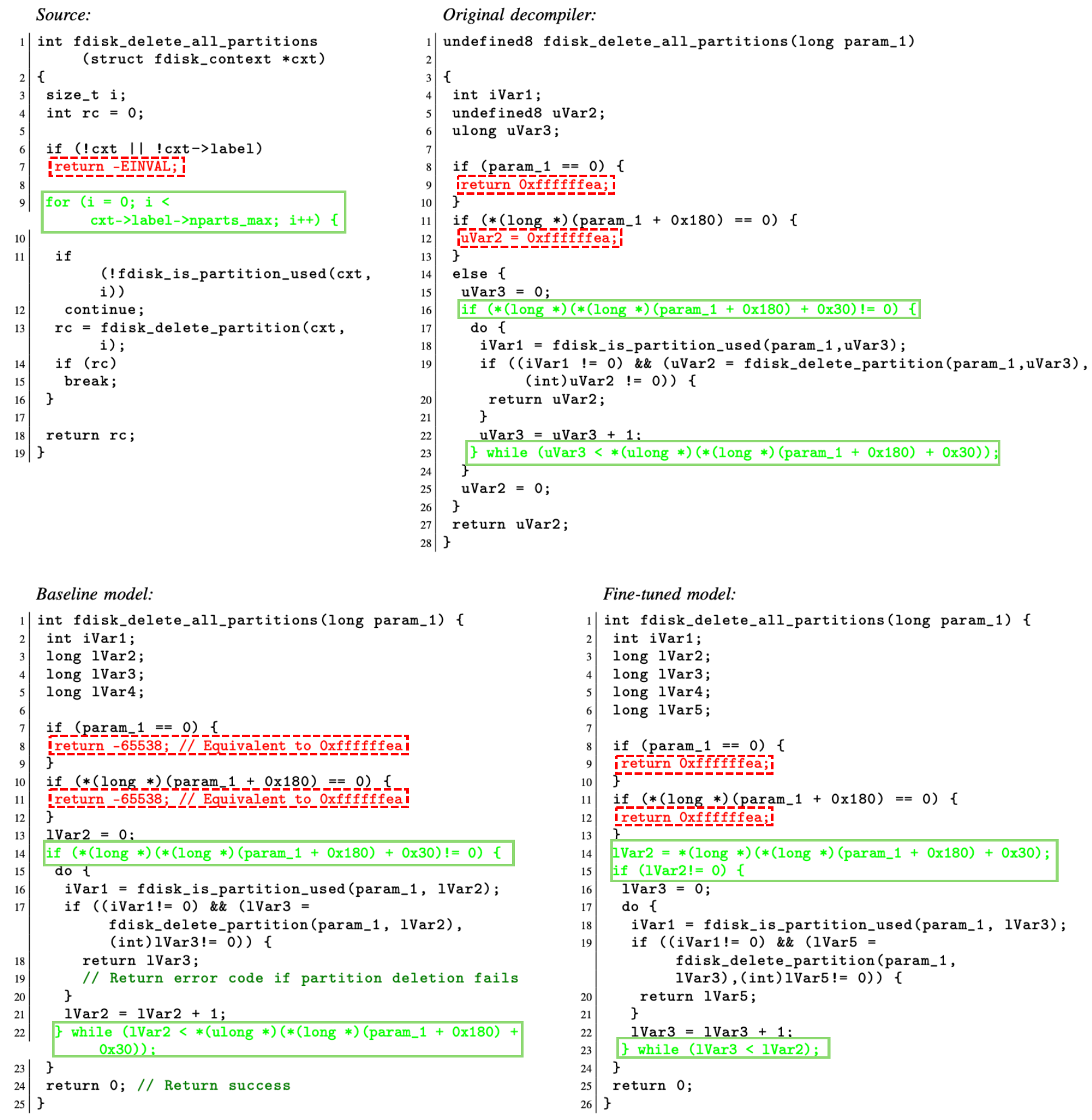}
\caption{This code snippet shows how \sys helps \texttt{Qwen2.5-Coder-3B} enhance its ability to generate more readable code (boxes with solid lines) while correcting the semantic error (boxes with dashed lines).}
    \label{fig:casestudy5}
\vspace{-0.75em}
\end{figure*}
\label{semanticcase}
\textbf{\textit{Fixing Incorrect Literal Value and Extracting Functional Variable.}} \Cref{fig:casestudy5} shows how the decompiled code snippet from the 
\\
\texttt{fdisk\_delete\_all\_partitions}  (\texttt{util-linux}) is improved by fixing the semantic errors while enhancing the readability.

Semantically, the baseline \texttt{Qwen2.5-Coder-3B} output mistakenly returns the decimal constant \texttt{-65538} (lines 8 and 11), which equates to \texttt{0xfffefffe} rather than the intended \texttt{0xffffffea}. 
Our fine-tuned model fixes this by emitting \texttt{return 0xffffffea;} at lines 9 and 12, restoring correct behavior.
For readability, our fine-tuned model makes two key improvements. 
First, it removes a redundant \texttt{if-else} construct  (original decompiled code line 14) and replaces it with a direct \texttt{return 0xffffffea} without an \texttt{else} condition added (fine-tuned model output line 12), yielding a control flow structure that more closely mirrors the original source. 
Second, it extracts the repeated expression \texttt{*(long *)(*(long *)(param\_1 + 0x180) + 0x30)} (fine-tuned model output line 14), into a named variable, reducing duplicated usage (original decompiled code line 15 and line 23).
As a result, \sys raises this function’s \syss from –2.000 to +0.625.
\newpage 
\clearpage

\bibliography{paper}


\end{document}